 \documentclass[journal,oneside]{IEEEtran}

\usepackage{ifpdf}

\usepackage{cite}

\ifCLASSINFOpdf
   \usepackage[pdftex]{graphicx}
\else
\fi
\usepackage{amsmath}
\usepackage{array}
\usepackage{balance}
\usepackage{color}

\begin{document}
%
\title{Resonant Scattering Characteristics of Homogeneous Dielectric Sphere}


\author{\IEEEauthorblockN{Dimitrios C. Tzarouchis,
Pasi Yl\"{a}-Oijala, 
and Ari Sihvola}
\thanks{This work is supported by the Aalto Energy Efficiency research program (EXPECTS project) and the ELEC Doctoral School Scholarship. 
The authors are with the Department of Electronics and Nanoengineering (formerly known as Radio Science and Engineering), Aalto University, 02150, Espoo, Finland. (dimitrios.tzarouchis@aalto.fi)}}

%



\IEEEtitleabstractindextext{%
\begin{abstract}
In the present article the classical problem of electromagnetic scattering by a single homogeneous sphere is revisited. Main focus is the study of the scattering behavior as a function of the material contrast and the size parameters for all electric and magnetic resonances of a dielectric sphere. Specifically, the Pad\'e approximants are introduced and utilized as an alternative system expansion of the Mie coefficients. Low order Pad\'e approximants can give compact and physically insightful expressions for the scattering system and the enabled dynamic mechanisms. Higher order approximants are used for predicting accurately the resonant pole spectrum. These results are summarized into general pole formulae, covering up to fifth order magnetic and forth order electric resonances of a small dielectric sphere. Additionally, the connection between the radiative damping process and the resonant linewidth is investigated. The results obtained reveal the fundamental connection of the radiative damping mechanism with the maximum width occurring for each resonance. Finally, the suggested system ansatz is used for studying the resonant absorption maximum through a circuit-inspired perspective.
\end{abstract}

}

\maketitle

\IEEEdisplaynontitleabstractindextext

%
\IEEEpeerreviewmaketitle

\section{Introduction}
%
%
%
%
\IEEEPARstart{A}{n} over-centennial and, perhaps, one of the most studied canonical problem encountered in electromagnetics is the scattering of electromagnetic radiation by a single sphere~\cite{Kuznetsov2016,Shore2015,Fan2014}. Its importance still resonates between classical and modern disciplines, such as RF engineering and nanotechnology, continuously delivering novel engineering achievements for a broad range of applications~\cite{Kuznetsov2016,Jahani2016,Tribelsky2015}. This problem is widely used either as a testbed for different material and morphological induced effects~\cite{Kuznetsov2016,Jahani2016,Lannebere2015} or for expanding our physical intuition about scattering aspects, such as non-radiating states, Fano resonances, and anomalous scattering, naming only a few~\cite{Kuznetsov2016,Lukyanchuk2010,Tribelsky2006}. All of the aforementioned occurring aspects become visible by the careful study and interpretation of the scattering processes on a simple sphere.

The electromagnetic resonant perspectives of a sphere can be described as a system for which the physical processes are encapsulated and described by the Mie coefficients. In order to reach a more complete picture regarding the resonant perspectives of a relatively small, homogeneous sphere, we revisit this problem, utilizing an alternative system ansatz for the Mie coefficients based on the recently introduced Pad\'e approximant technique~\cite{baker1996pade,Tzarouchis2016c}. 

The aim of this article is to propose and discuss a systematic way for studying the fundamental resonant scattering aspects caused by a subwavelength sphere. We concentrate our main efforts in exposing the resonant conditions and their size dependencies of the lowest multipole (dipole, quadrupole, etc.) magnetic and electric resonances. In parallel, the effects of the radiative damping mechanism on the scattering process are revealed, especially about the resonant width and the absorption maximum of each mode. These, non-trivial, aspects reinforce our understanding regarding the light scattering/absorptive characteristics of a sphere, being readily expandable for other canonical-shaped geometries.

This work is divided into four general parts. Section~\ref{sec:mie} discusses some observations derived directly from the analysis of the Mie coefficients. Section~\ref{sec:compact} presents compact expansions of the main coefficients in an attempt to extract physical intuition through the Pad\'e approximants.
Section~\ref{formulas} discusses more accurate pole conditions, which utilize higher order Pad\'e expansions; in addition, the accuracy of the proposed pole conditions and the widths of the Mie resonances are discussed. Finally, Section~\ref{sec:abs} presents a circuit-inspired analysis, revealing the importance of the radiative damping term to the overall scattering process. Conclusions and appendices complete the discussion about the proposed perspective. The reader who wishes to extract only the analytical pole conditions formulas regarding the electric and magnetic resonances of a dielectric sphere can skip the analysis sections and go directly to the tables provided in Section~\ref{formulas}.

\section{Mie--scattering Preliminaries}\label{sec:mie}
Let us begin by briefly introducing the mathematical model describing the electromagnetic scattering by a small homogeneous sphere of radius $a$ illuminated by a monochromatic plane wave ($e^{-i\omega t}$); an extensive review can be found in~\cite{Shore2015}. It is well known that the scattered fields can be decomposed into set of spherical harmonics, whose complex amplitudes are given by the $a_n$ and $b_n$ Lorenz--Mie coefficients~\cite{bohren2008absorption} viz.,
\begin{equation}\label{an}
 a_n=\frac{m^2 j_n(mx)\left[xj_n(x)\right]'-\mu_c j_n(x)\left[mxj_n(mx)\right]'}{m^2 j_n(mx)\left[xh_n^{(1)}(x)\right]'-\mu_c h_n^{(1)}(x)\left[mxj_n(mx)\right]'}
\end{equation}

\begin{equation}\label{bn}
 b_n=\frac{\mu_c j_n(mx)\left[xj_n(x)\right]'-j_n(x)\left[mxj_n(mx)\right]'}{\mu_c j_n(mx)\left[xh_n^{(1)}(x)\right]'-h_n^{(1)}(x)\left[mxj_n(mx)\right]'}
\end{equation}
where $x=ka$ is the free space size parameter, $m=\frac{k_c}{k}=\sqrt{\varepsilon_c\mu_c}$ is the material contrast, and $\varepsilon_c=\varepsilon_1/\varepsilon$ and $\mu_c=\mu_1/\mu$ are the permittivity and permeability contrasts, respectively.

These $a_n$ and $b_n$ terms are generally complex and demonstrate the electric and magnetic multipole contributions to the overall scattering process~\cite{stratton2007electromagnetic}. Usually, most of the far field scattering processes are quantified through these two coefficients. Note that these coefficients can be found in the literature by many names, i.e., TM and TE modes, E-wave and H-wave coefficients~\cite{bohren2008absorption,stratton2007electromagnetic} etc.. Here we will refer to them as electric ($a_n$) and magnetic ($b_n$) multipole terms, having for each value of $n=1,2,3,...$ names such as dipole ($n=1$), quadrupole ($n=2$) and so on. Similarly, the internal fields can be quantified through the $c_n$ and $d_n$ coefficients, viz.,
\begin{equation}\label{dn}
 d_n=\frac{\mu_c m j_n(x)\left[xh^{(1)}_n(x)\right]'-\mu_c m h^{(1)}_n(x)\left[xj_n(x)\right]'}{m^2 j_n(mx)\left[xh_n^{(1)}(x)\right]'-\mu_c h_n^{(1)}(x)\left[mxj_n(mx)\right]'}
\end{equation}

\begin{equation}\label{cn}
 c_n=\frac{\mu_c j_n(x)\left[xh^{(1)}_n(x)\right]'-\mu_c h^{(1)}_n(x)\left[xj_n(x)\right]'}{\mu_c j_n(mx)\left[xh_n^{(1)}(x)\right]'-h_n^{(1)}(x)\left[mxj_n(mx)\right]'}
\end{equation}
where $d_n$ and $c_n$ corresponds to the electric and the magnetic multipole term, respectively. 

All coefficients, i.e., Eqs.~(\ref{an}),~(\ref{bn}),~(\ref{dn}), and~(\ref{cn}), can rigorously describe the material and size dependences of the scattering and internal fields. However their complex structure prevent us from obtaining a simple physical interpretation on their resonant behavior. Therefore, a question arises: how can we extract any physically intuitive information regarding their size- and material-dependent characteristics?

For the quasistatic limit ($x\rightarrow0$) these terms are usually expanded in a Taylor (Maclaurin) series~\cite{stratton2007electromagnetic}, revealing some interesting perspectives e.g, the Rayleigh scattering limit~\cite{bohren2008absorption}. The series expansion of Eqs.~(\ref{an}) and~(\ref{bn}) for small $x$ read 
\begin{equation}\label{a1Taylor}
 a_1^T\approx-i\frac{2}{3}\frac{\varepsilon_c - 1}{\varepsilon_c +2 }x^3-i\frac{1}{5}\frac{\varepsilon_c^2(1+\mu_c)-6\varepsilon_c+4}{(\varepsilon_c +2 )^2}x^5+O[x]^6
\end{equation}

\begin{equation}\label{b1Taylor}
 b_1^T\approx-i\frac{2}{3}\frac{\mu_c - 1}{\mu_c+2}x^3-i\frac{1}{5}\frac{\mu_c^2(1+\varepsilon_c)-6\mu_c+4}{(\mu_c+2)^2}x^5+O[x]^6
\end{equation}

\begin{equation}\label{d1Taylor}
 d_1^T\approx\frac{3}{\epsilon_c+2}+\frac{3}{10}\frac{\left(\mu_c\left(\epsilon_c^2+4\epsilon_c\right)+5\epsilon_c-10\right)}{(\epsilon_c+2)^2}x^2 +O\left[x\right]^{5/2}
\end{equation}

\begin{equation}\label{c1Taylor}
 c_1^T\approx\frac{3 \mu_c}{(\mu_c+2) \sqrt{\mu_c \epsilon_c}}+O\left[x\right]^{3/2}
\end{equation}

Indeed, many intuitive results are visible in this simple system expansion, such as the electrostatic polarization enhancement condition ($\varepsilon_c=-2$), also known as Fr\"{o}hlich frequency~\cite{bohren2008absorption,sihvola2006character}. The external coefficients suggests a dual behavior, i.e., invariance with respect to interchanging $\varepsilon_c$ and $\mu_c$. However, the same does not apply directly for the internal ones. For instance, the $c_n$ coefficient possesses two resonant conditions, $\mu_c=-2$ and $\varepsilon_c=0$, while $d_n$ only for $\varepsilon_c=-2$.

\begin{figure}[!]
\centering
 \includegraphics[width=0.5\textwidth]{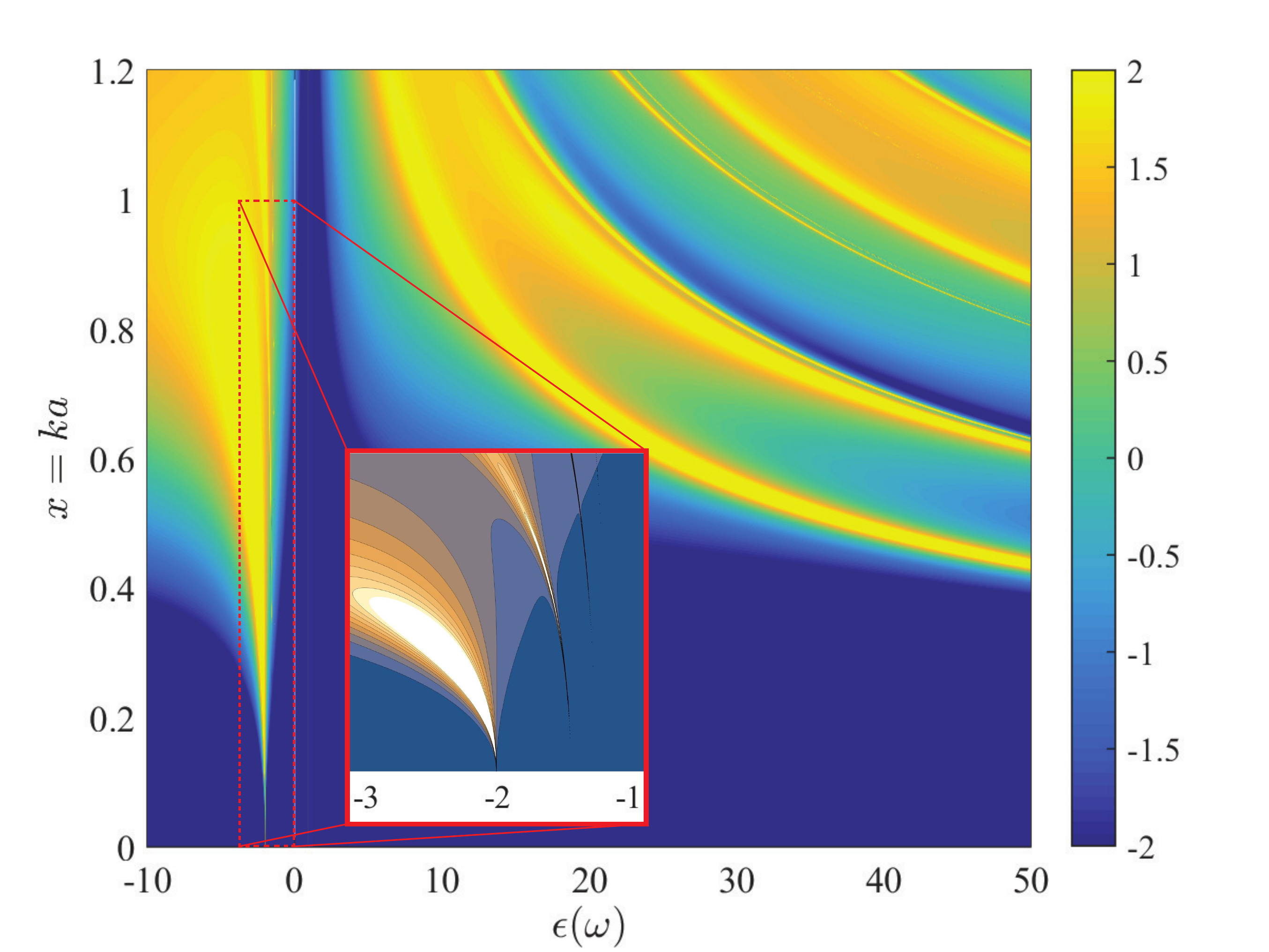}
         \caption{The scattering efficiency of a lossless dielectric sphere, as a function of the material permittivity and the size parameter. There are two visible regions: the plasmonic region for $\varepsilon<0$ and the dielectric region ($\varepsilon_c>0$). In both regions vibrant resonances occur. The  dielectric resonances are geometrical induced resonances, of either magnetic or electric origin, respectively. The plasmonic resonances are due to the free oscillating charges, thus characterized as electric. The inset image depicts (only a few of) the main plasmonic resonances residing in region $-3<\varepsilon(\omega)<-1$.}
        \label{Qext}
\end{figure}

In this work we focus on the resonant conditions and the physical mechanisms for a small, magnetically inert ($\mu_c=1$), homogeneous sphere. Fig.~\ref{Qext} depicts the resonant scattering extinction spectrum of a lossless sphere as a function of the material and size parameters. Several resonances occur for both positive and negative permittivity values. Note that the scattering efficiency depends on $a_n$ and $b_n$, i.e.,  $\sigma_{ext}\propto\sum^\infty_{n=1}\left(\Re\{a_n\}+\Re\{b_n\}\right)$. Based on the values of permittivity we categorize the resonances as plasmonic ($\varepsilon(\omega)<0$), and dielectric ($\varepsilon(\omega)>0$). As we can see in Fig.~\ref{Qext}, the scattering spectrum reveals a wealth of resonances in both regions with several qualitative differences. For example the plasmonic resonances may appear even for very small spheres.

\begin{figure}[!]
\centering
 \includegraphics[width=0.45\textwidth]{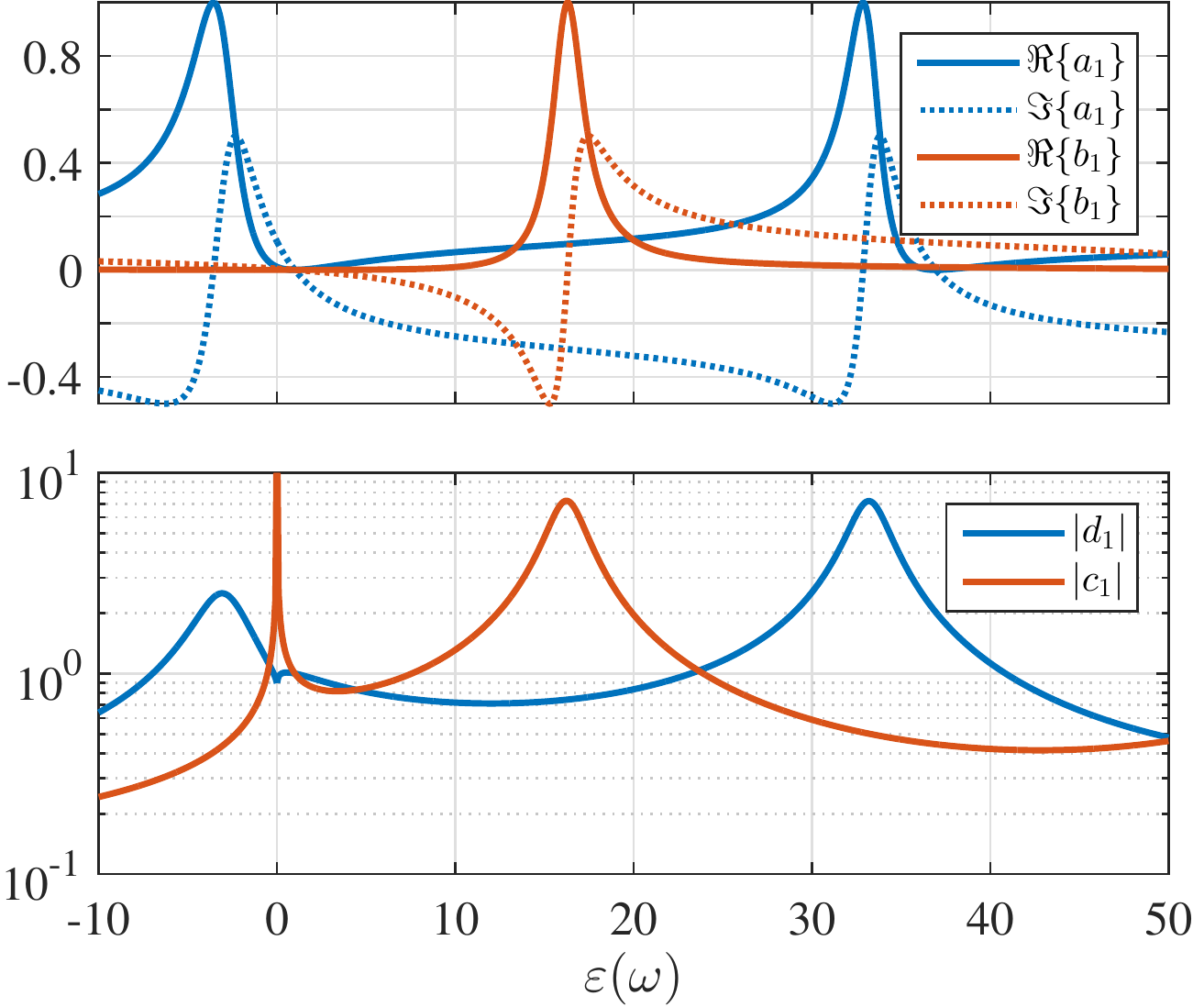}
         \caption{The external scattering coefficients $a_1$ and $b_1$ (up), and the internal coefficients $d_1$ and $c_1$ for $x=0.75$ as a function of the permittivity (lossless case). The absolute value of the external coefficients cannot exceed unity due to passivity while the amplitude of the internal coefficients is not bounded.}
        \label{coeff}
\end{figure}

As a reference, Fig.~\ref{coeff} depicts the real and imaginary values of both external and internal Mie coefficients for a given size parameter ($x=0.75$). One observes that both external and internal coefficients exhibit maximum values in both negative and positive permittivity regions. Although the external coefficients are bounded, i.e., they reach a maximum unity amplitude for the lossless case, the internal coefficient are unbounded to a specific value. Hence they may exhibit a qualitatively different resonant lineshape than the external ones.

A usual treatment for extracting the resonant conditions is derived by observing that Eq.~(\ref{an}) and~(\ref{bn}) can be decomposed in a rational form $\frac{r_n}{r_n+is_n}$, where $r_n$, $s_n$ are real functions for the lossless case~\cite{Ruppin1975}. The approach in~\cite{Tribelsky2006,Osipov2015} is that each $r_n$ and $s_n$ term is expanded in a Taylor series, enabling in such a way a pole condition study. However, the Taylor expansions might suffer from a slow convergence, especially near the poles of the system. Similar rational expansions were given in~\cite{Wiscombe1980} for a quick numerical evaluation of the Mie coefficients for small spheres. Furthermore, for the case of optically large spheres several asymptotic formulas have been also derived in~\cite{Lam1992}.

In order to find a suitable coefficient expansion with which to unlock the material and size dependencies we utilize an alternative approximative perspective based on the Pad\'e approximants for the Mie coefficients, recently introduced in~\cite{Tzarouchis2016c}. Earlier, Pad\'e-like expansions have been given for certain of the scattering characteristics, e.g., fixed material parameters~\cite{Conwell1984} or small spheres~\cite{Alam2006,Ambrosio2012}. There, however, the focus was not in identifying neither the resonant conditions nor the physical processes. 

The Pad\'{e} approximants are a special type of rational (approximative) expressions constructed by expressing a function in a rational form and expanding each of the rational terms in power series~\cite{baker1996pade}. Here we follow the notation $[L/M]$ for the Pad\'e expansion, corresponding to a fractional expression of two polynomials $\frac{P(x)}{Q(x)}$ of order $x^L$ and $x^M$, respectively. This polynomial formalism allows us to approximate a given function as a set of zeros and poles. Thus, this type of approximation is particularly useful for functions or systems whose physical responses include resonances throughout their spectrum, and is to be preferred from Taylor series expansion due to slow convergence of the latter close to pole values~\cite{baker1996pade}. %

\section{Compact Expansions for the Mie coefficients: extracting physical intuition}\label{sec:compact}

We start by giving the first $[3/3]$ Pad\'e expansion for $a_1$

\begin{equation}\label{a133}
 a_1^P\approx-i\frac{2}{3}\frac{\varepsilon_c-1}{\varepsilon_c+2}\frac{x^3}{\left(1-\frac{3}{5}\frac{\varepsilon_c-2}{\varepsilon_c+2}x^2-i\frac{2}{3}\frac{\varepsilon_c-1}{\varepsilon_c+2}x^3\right)}
\end{equation}
exhibiting a complex pole of the form
\begin{equation}\label{a1_pole_1}
\varepsilon_c=-2\frac{15i+9ix^2-5x^3}{15i-9ix^2+10x^3}
\end{equation}
Equation~(\ref{a133}) can be decomposed into three distinctive parts. First, the $\frac{\varepsilon_c-1}{\varepsilon_c+2}$ part can be recognized as the \emph{static} (Rayleigh) term, similarly with Eq.~(\ref{a1Taylor}). This term contributes a system pole at $\varepsilon_c=-2$ only for the static case~\cite{Mei2013}. The second part is the $\frac{\varepsilon_c-2}{\varepsilon_c+2}x^2$ term found in the denominator of Eq.~(\ref{a133}). This term introduces the size effects over the coefficient and to the overall scattering procedure, and thus it is recognized as the \emph{dynamic depolarization} term~\cite{Meier1983}. Lastly, the imaginary term of $x^3$ order is recognized as the \emph{radiative damping} part. For the lossless case ($\varepsilon''=0$) this part prevents the coefficient from infinite growth and represents the intrinsic radiative damping/reaction mechanism of the scattering process~\cite{DeVries1998}.

By expanding the pole condition~(\ref{a1_pole_1}) up to the fifth order, we obtain
\begin{equation}\label{a1_pole_2}
\varepsilon_c=-2-\frac{12}{5}x^2(1+\frac{3}{5}x^2)-2ix^3(1+\frac{7}{5}x^2)+...
\end{equation}
In a similar manner each term to the pole condition can be recognized, i.e., static ($-2$), dynamic ($-\frac{12}{5}x^2$), and damping term ($-2ix^3$), respectively. Notice that a similar but somewhat less accurate condition can be readily found in~\cite[Ch.12, p.329]{bohren2008absorption}.

As for the magnetic coefficient, the $[5/5]$ Pad\'e $b_1$ expansion is 
\begin{equation}\label{b155}
 b_1^P\approx-i\frac{\varepsilon_c-1}{45}\frac{x^5}{\left(1+\frac{1}{21}(5-2\varepsilon_c)x^2+[x^4]-i\frac{1}{45}(\varepsilon_c-1)x^5\right)}
\end{equation} 
with the truncated term to be
\begin{equation}
 [x^4]=-\frac{\varepsilon_c^2+100\varepsilon_c-125}{2205}x^4
\end{equation}
Expression~(\ref{b155}) gives a resonant condition yielding to the value

\begin{equation}\label{b1_small}
\varepsilon_c=-2.07+\frac{10.02}{x^2}+1.42 x^2-2 i x \left(1.06-0.77x^2\right)
\end{equation}
with accuracy up to the second decimal. However, not every pole of the Pad\'{e} expansions correspond to an observed pole~\cite{baker1996pade}. This is happening for three reasons. The Mie terms are described by a set of complex functions, hence all the poles exist \emph{a-priori} in the complex plane. These complex poles, known also as \emph{natural frequencies}~\cite{stratton2007electromagnetic}, may exhibit large imaginary parts. Thus their physical observation is extremely difficult. Secondly, the Pad\'e approximants can give some intrinsic, non-existent poles~\cite{Masjuan2014} to the system. Finally, some of the poles can be mutually canceled by system's zeros, and therefore cannot be observed. For the above reasons we will restrict our analysis only to physically observable poles, verified by the Mie spectrum.

The condition described in Eq.~(\ref{b1_small}) reveals certain interesting phenomena regarding the nature of the magnetic resonances enabled on a dielectric sphere. First, there is an inverse square dependence, showing that the magnetic resonance for small spheres can be reached only for huge contrast materials, making very difficult its observation for very small spheres. Secondly, there is a constant term, slightly regulating the real part of the pole condition. Lastly, the radiative damping process of a magnetic dipole does not follow the same volume dependence ($x^3$) observed in the previous electric dipole case (Eq.~(\ref{a1_pole_2})), but rather a linear $x$ dependence. Actually, a closer look reveals that this dependence in not exactly linear, exposing a plateau for size parameters around $0.5$~\cite{Tzarouchis2016c}. 

The Pad\'e expansions $[0/3]$ and $[0/5]$ of the internal coefficients read
\begin{equation}\label{d1P}
 d_1^P\approx\frac{3}{(\varepsilon_c+2) \left(1-\frac{1}{10}\frac{\varepsilon_c^2+9 \varepsilon_c-10}{\varepsilon_c+2}x^2-i\frac{2}{3}\frac{\varepsilon_c-1}{\varepsilon_c+2}x^3\right)}
\end{equation}

\begin{equation}\label{c1P}
 c_1^P \approx\frac{1}{\sqrt{\varepsilon_c} \left(1+\frac{1}{6} (1-\varepsilon_c) x^2+[x^4]-i\frac{1}{45}  (\varepsilon_c-1) x^5\right)}
\end{equation}
where $[x^4]=\frac{1}{120} \left(\varepsilon_c^2-6 \varepsilon_c+5\right) x^4$. Following the same pole analysis as above we obtain two resonances for the $d_1$ coefficient, one of which follows exactly the value of Eq.~(\ref{a1_pole_2}) for the external $a_1$ coefficient, while the other reads
\begin{equation}\label{elec_pole}
 \varepsilon_c=-7+\frac{10.}{x^2}
\end{equation}
This condition is a very rough approximation for the electric resonances for $\varepsilon_c>0$. The accurate evaluation of this pole requires a higher order Pad\'e expansion and its form will be extracted in Section~\ref{formulas}. Notice that also this pole exhibits an inverse square size resonant condition, similar to the magnetic resonances. 

Expression~(\ref{c1P}) gives three poles, two of which are spurious, while the remaining one is at $\varepsilon_c=0$. This pole condition exists for every $c_n$ coefficient, since in every expansion there is a term of the form $\frac{1}{\varepsilon_c^{n/2}}$, where $n$ is the order of the coefficient. In other words the $c_n$ coefficient experiences a epsilon-near-zero (ENZ) resonance for every multipole term. ENZ behavior is closely related to what it is called perfect magnetic conductor (PMC), and it is conceptually used for light trapping~\cite{Lannebere2015} and shape-invariant resonant cavities~\cite{Liberal2016}. Surprisingly, this is not true for the mu-near-zero case, since the $d_n$ coefficients are not dual to $c_n$ in terms of the material parameters ($a_n$ and $b_n$ are dual, as seen from the expansion Eqs.~(\ref{a1Taylor}) and~(\ref{b1Taylor})). 
\begin{figure}[!]
\centering
 \includegraphics[width=0.45\textwidth]{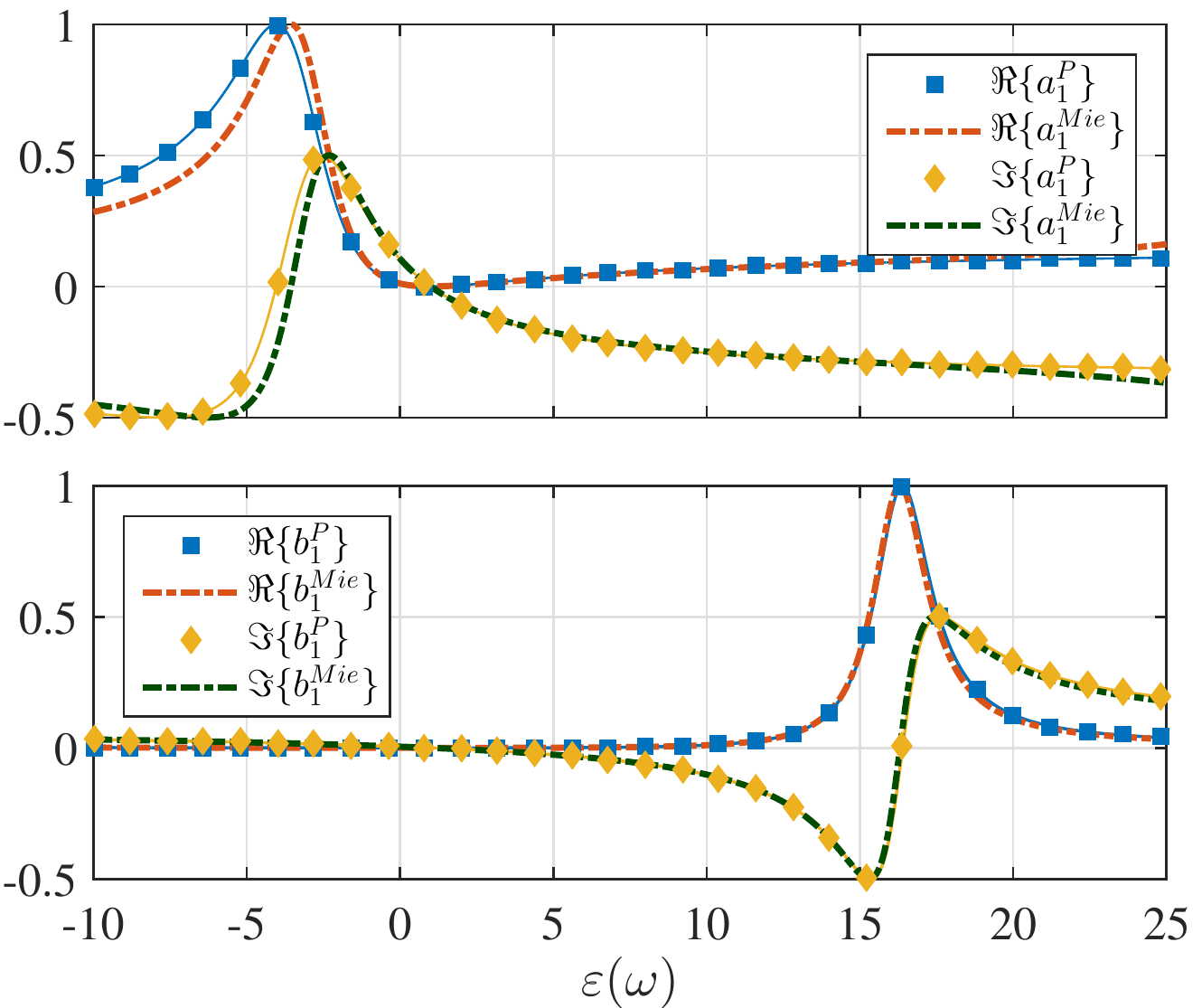}
         \caption{The real and imaginary parts of the first dipole coefficients ($a_1$--top, $b_1$--bottom) for a sphere of $x=0.75$ as a function of the permittivity. Top: The blue-cube and yellow-diamond lines depict the real and imaginary part of the $[3/3]$ Pad\'e expanded expression ($a^P_1$ in Eq.~(\ref{a133})) compared with the Mie values (red-dashed: real part, black-dashed: imaginary part) of Eq~(\ref{an}). Bottom: The blue-cube and yellow-diamond lines here depict the real and imaginary part of the $[5/5]$ Pad\'e expanded expression of $b^P_1$ (Eq.~(\ref{b155})) compared with the Mie values (red-dashed real part, black-dashed imaginary part) of Eq~(\ref{bn}). Note that the position, the strength, and the linewidth of the curves coincide sufficiently, verifying the validity of the proposed system ansatz even for relatively large spheres.}
        \label{lineshape_comp}
\end{figure}

\subsection*{Electric resonances for $\varepsilon_c>0$}

The above analysis made clear that small-order Pad\'{e} expansions provide us with simple expressions for all  dipole $a_1$, $b_1$, $d_1$, and $c_1$ Mie coefficients and their corresponding pole conditions. The Pad\'{e} approximants are able to predict the zeros and the poles of this system. In order to obtain more system poles for $a_1$ in the $\varepsilon_c>0$ regime, higher order approximants should be used. However, the Pad\'e expressions of these expansions are lengthy, hence will be omitted.

To start with, the [7/2] expansion of the $a_1$ coefficient gives in total four poles, one corresponding to the plasmonic case, two non-observable double complex roots, and a fourth yielding the following expression
\begin{equation}\label{a1P+72}
 \varepsilon_{c}^{[7/2]}=-14.5+\frac{22.5}{x^2}+i\frac{77.778}{x}+...
\end{equation}
where the superscript indicates the order of the used Pad\'e approximant.

This first, very rough pole condition, exhibits the same inverse square size dependence observed in Eq.~(\ref{elec_pole}). Here the radiative damping term is extremely large, an indication that a more accurate, higher order expansion is needed. To do so we increase the order of the numerator, keeping the denominator order as low as possible. By this, straightforward, and numerically convenient heuristic method we are able to identify more accurately any of the pole conditions. After several iterations a quite accurate pole condition for the electric resonance occurs from a $[19/2]$ Pad\'e expansion, converging to the value 
\begin{equation}\label{a1P_192}
\begin{aligned}
 \varepsilon_{c}^{[19/2]}=&-1.99546+\frac{20.193}{x^2}-2.05137x^2\\
 &+i0.0241962x-i1.91773x^3+...
\end{aligned}
 \end{equation}

An interesting remark can be drawn here; the estimated radiative damping term (affecting the absorption maximum, see Section~\ref{sec:abs}) gives both an $x$ and $x^3$ dependence. However the first linear term is a numerical artifact, converging quickly to a close-to-zero value. The the $x^3$ term converges close to two, resembling the plasmonic radiative damping term. This is a reasonable result since both resonances are of electric dipole nature. Eq.~(\ref{a1P_192}) exhibits a linear imaginary term, two orders of magnitude smaller than the magnetic, demonstrating some fundamental differences between the electric and the magnetic dipole resonances on a sphere.

\subsection*{Resonant Lineshape}

By plotting the compact expressions given by Eqs.~(\ref{a133}) and~(\ref{b155}) versus the Mie coefficients $a_1$ and $b_1$ (Fig.~\ref{lineshape_comp}), one observes that even for relatively large spheres ($x=0.75$) the resonant lineshape almost coincides with the exact one obtained by the full Mie solution. The Pad\'e approximants can be used instead the complex Mie terms to emulate the exact physical behavior of the scattering sphere. This is an attractive feature where simple and physically intuitive models are needed, e.g, homogenization models for composite materials and surfaces~\cite{Jahani2016}.

\section{Pole conditions with increased accuracy}\label{formulas}

Throughout the aforementioned analysis it has become clear that low order Pad\'e approximants can be used for extracting intuition regarding the physical mechanisms of the system. By using higher order approximants and applying the same heuristic method described above, more accurate pole conditions can be extracted for all resonances. However, these approximants are lengthy and will be omitted. Details about the pole conditions can be found in Appendix~\ref{Ap_b}. Table~\ref{formulas_table} contains the converged formulas of the first resonances for the first five ($n=1,2,3,4,5$) magnetic $b_n$ coefficients, and the first four electric $a_n$ coefficients for the dielectric case with parametric values listed in Table~\ref{values}.

The above generalized conditions give the value of the complex permittivity value required for the system to resonate. These expressions can be readily used as design rule-of-thumb equations. The pole conditions of the electric resonances, found in Table~\ref{formulas_table}, second equation and Eq.~(\ref{general_an2}), reveal a similar but not identical imaginary part; only the $a_1$ (dipole resonance) exhibits the same $x^3$ imaginary part dependency. This minor point describes that the behavior of the electric resonances in the plasmonic and dielectric region is close but not completely identical. Note that the $p_n$ coefficients correspond to the first root of the $n$-th spherical Bessel function of the first kind ($j_n(p_n)=0$). The dots in Table~\ref{values} denote that the higher order $t_n$ values can be extracted in a similar way for resonances above $n=4$.

\begin{table}[h]
\centering
\caption{Pole formulas of the dielectric resonances for $n=1,2,3,...$}
\label{formulas_table}
\begin{tabular}{l}
\hline
\\
$\varepsilon_{bn}=-\frac{2}{2n-1}+\left(\frac{p_n}{x}\right)^2-i\frac{2}{\left[\left(2n-1\right)!!\right]^2}x^{2n-1}\left(1-t_nx^2\right)
$
\\
\\
$
\varepsilon^{diel}_{an}=-\frac{2}{n}+\left(\frac{p_{n+1}}{x}\right)^2-i\frac{2}{\left[n\left(2n-1\right)!!\right]^2}x^{2n+1}\left(1-\frac{t'_{n+1}}{x^2}\right)
$

\end{tabular}
\end{table}

\begin{table}[h]
\begin{center}
\caption{Values of uses parameters for the $b_n$ and $a_n$ pole resonances}
\begin{tabular}{l*{5}{c}{c}}\label{values}
$n$              & 1 & 2 & 3 & 4 & 5 \\
\hline\hline
$p_n$ & $\pi$ & 4.4934 & 5.7634 & 6.9879 & 8.1428 \\ 
$t_n$            & 0.6960 & 0.2508 & 0.1560 & 0.1166 & ... \\ 
$t'_n$           & - & 0.0040 & 0.0954 & 0.3461 & 1.3923  \\
\end{tabular}
\end{center}

\end{table}
Finally, a general resonant condition for the plasmonic resonances ($\varepsilon(\omega)<0$) reads~\cite{Tribelsky2011}
\begin{equation}\label{general_an2}
\begin{aligned}
 \varepsilon^{plasm}_{an}=&-\frac{n+1}{n}+\frac{2(2n+1)(n+1)}{n^2(2n-1)(2n+3)}x^2\\
 &-i\frac{n+1}{\left[n\left(2n-1\right)!!\right]^2}x^{2n+1},~n=1,2,...
\end{aligned}
\end{equation}

\subsection*{Error analysis}

\begin{figure}[!]
\centering
 \includegraphics[width=0.5\textwidth]{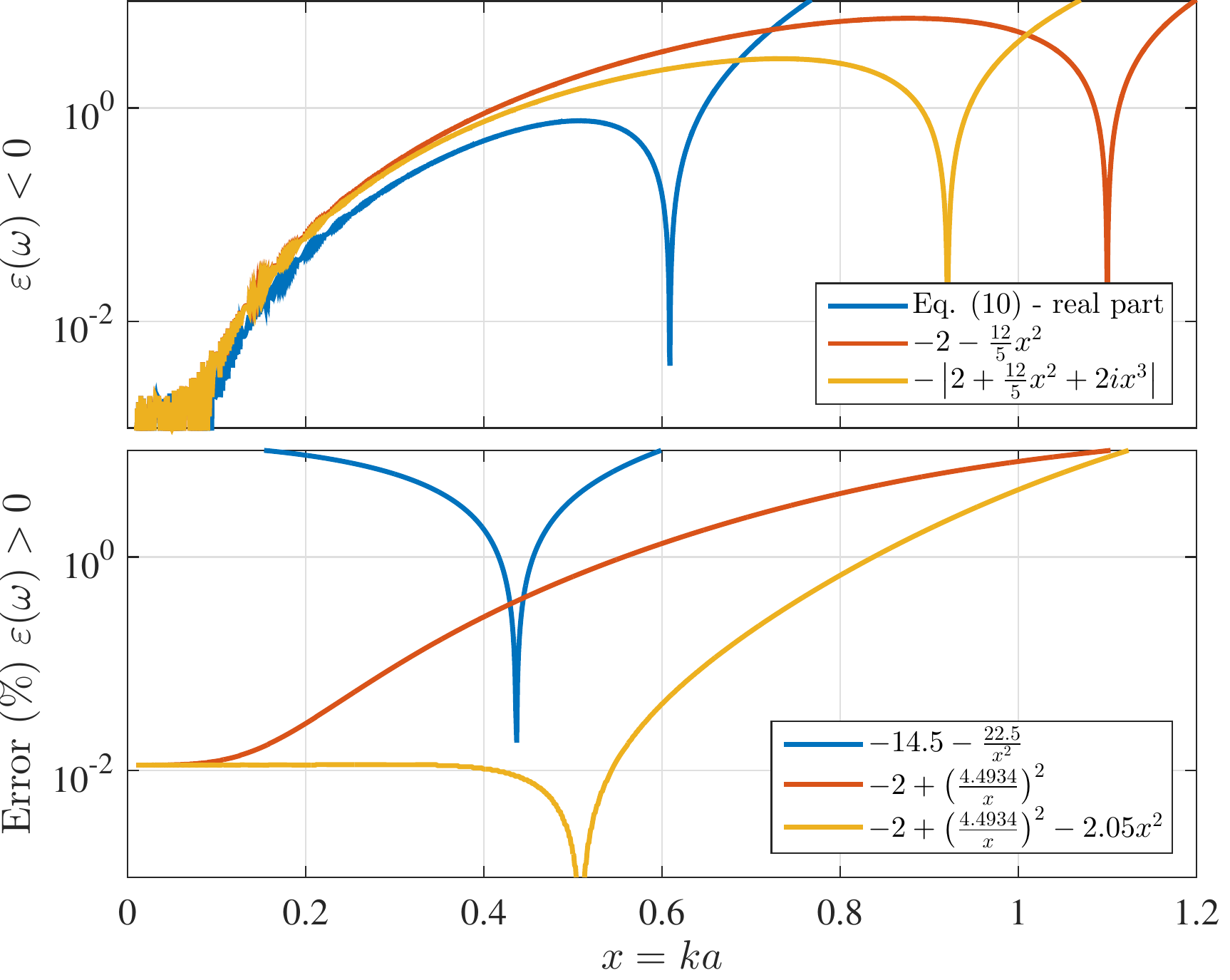}
         \caption{The relative error (in percentage) between the actual pole position of the first electric resonance ($a_1$) as a function of the size parameter. Top: Error for the plasmonic resonance. Bottom: Error for the dielectric resonances.}
        \label{error}
\end{figure}

The accuracy of the obtained formulas with respect to the full Mie resonant position is depicted in Fig.~\ref{error}, for the first electric resonances in the plasmonic and dielectric case. Specifically, an pole condition with more than $1\%$ accuracy can be achieved up to $x\approx0.6$ for the plasmonic case. This value is obtained by considering only the real part of Eq.~(\ref{general_an2}). By introducing the imaginary part and evaluating the absolute value of Eq.~(\ref{general_an2}) the accuracy is increased; the error reach a $3\%$ value for $x\approx0.7$ and drops, expanding the accurate region up to $x\approx0.95$. 

For the the dielectric case the second order term gives accurate predictions for sizes up to $x\approx0.8$. Similar results regarding the magnetic coefficient can be found in~\cite{Tzarouchis2016c}, where the pole condition (Table~\ref{formulas_table}, first equation) exhibits less than $1\%$ the error for size parameters up to $0.6$.

\subsection*{Width of the resonances}

The proposed Pad\'e expansion can also provide information about the linewidth of the resonances. As we can see in Fig.~\ref{lineshape_comp} the approximated line shape can be very close to the analytical one even for relatively large particles, i.e., up to $x=0.75$. This feature can be particularly useful for evaluating the linewidth of the scattering process. It is known that the full width half maximum (FWHM) scattering linewidth, i.e., $|a_1|^2=\frac{1}{2}$ for the electric dipole, is restricted by the intrinsic material losses~\cite{Johnson1993,Averitt1999}.

The partial extinction efficiency of each electric and magnetic term is proportional to the real part of each coefficient, i.e., $\sigma^{a_n}_{ext}\propto\Re\{a_n\}$ and $\sigma^{b_n}_{ext}\propto\Re\{b_n\}$. Let us simplify the analysis by considering only the first electric extinction cross section. For the lossless case the extinction linewidth reads 
\begin{equation}\label{BW}
 \Delta\varepsilon(\omega)=4x^3
\end{equation}
revealing that the linewidth is twice the radiative damping term, explaining in a way the fact that radiative damping affects the scattering strength, linewidth and, consequently, the radiative decay rates~\cite{Carminati2006}. This result can be generalized for every multipole term through 
the imaginary part of conditions in Table~\ref{formulas_table} and~(\ref{general_an2}), respectively. For example the width of the first three plasmonic resonances is of the order $x^3$, $x^5$, and $x^7$, respectively. Note that  higher order modes experience a very narrow bandwidth, and hence are very difficult to be tracked, as also seen in Fig.~\ref{Qext}.

The previous result has a fundamental connection with the bandwidth bounds derived for an electrically small resonator, namely with the quality factor of an electrically small antenna (ESA), expressed as $Q_{\text{ESA}}\approx\frac{1}{x^3}+\frac{1}{x}$~\cite{Collin1964}. This bound is derived from the decomposition of the antenna secondary field into a set of spherical harmonics~\cite{jackson1999classical,Gustafsson2012,Hansen2012}. In our analysis, the linewidth (BW) of a very small plasmonic scatterer is of the order of $x^3$, since only the electric dipole term is present. Similarly, for a small dielectric resonator (magnetic dipole) the linewidth is $x$. Assuming $Q\approx$ BW, the $Q_{\text{ESA}}$ value can be reached qualitatively from the combination of the first electric and the first magnetic resonance. Consequently, the decomposition of the secondary scattering field by an arbitrary shaped scatterer into spherical harmonics can give a straightforward hint about its resonant bandwidth.

Let us now assume material losses, i.e., $\varepsilon(\omega)=\varepsilon^\prime(\omega)+i\varepsilon^{\prime\prime}(\omega)$. In this case the extinction linewidth is a function of both real and imaginary part of permittivity. The maximum value of the extinction cross section as a function of $\varepsilon'$ can be found by evaluating the position on which the first partial derivative is zero, i.e., $\frac{\partial \sigma(\varepsilon',\varepsilon'')}{\partial \varepsilon'}=0$. This procedure gives that the resonant linewidth, with respect to the $\varepsilon'$ is 
\begin{equation}\label{BW_loss}
 \Delta\varepsilon'(\omega)=2\varepsilon''+4x^3
\end{equation}
demonstrating that total linewidth is a superposition between the material losses and the radiative damping. Obviously for zero material losses Eq.~(\ref{BW_loss}) yields to the lossless case of Eq.~(\ref{BW}). 

These material-model independent results demonstrate that not only the material losses but the radiative damping mechanism is responsible of the width of the scattering spectrum. Note that the absorptive linewidth exhibits a more complex behavior as a function of the material losses. Hence, in the following section we will study how the scattering and material losses affect the overall absorptive characteristics.

\section{Circuit-inspired analysis of the scattering process: maximizing absorption efficiency}\label{sec:abs}

To this point, the Pad\'{e}-expanded Mie coefficients exhibit reduced complexity, revealing a set qualitative characteristics about the underlying physical mechanisms. In other words we exchanged mathematical exactness for physical intuition. In this section the aim is to exploit further the Pad\'e expansions for finding the necessary conditions that maximize the absorption efficiency of a small homogeneous sphere.

By carefully examining the approximated Mie coefficients, one can rewrite them in the following simple form e.g., 
\begin{equation}\label{eq:circuit}
 a_1^P=\frac{Z_R}{Z_R+Z_L}=\frac{1}{1+\frac{Z_L}{Z_R}}
\end{equation}
where the impedances are
\begin{equation}
 Z_R=2\varepsilon''x^3-2i(\varepsilon'-1)x^3=R_R+iX_R
\end{equation}
and
\begin{equation}
\begin{aligned} 
Z_L= & ~ 3(\varepsilon'+2)-\frac{9}{5}(\varepsilon'-2)x^2+i 3\varepsilon''\left(1-\frac{3}{5}x^2\right)\\
  = &  ~ R_L+iX_L
\end{aligned}
\end{equation}

Arguably, $Z_L$ possesses a non-passive character ($R_L<0$) for certain size and material combinations, a fact that has been identified in~\cite{Alu2011}. The quantity $\frac{Z_L}{Z_R}$ yields to the value of $iX_L$, known also as reactance~\cite{Ruppin1975}. In this section we try to extract the necessary conditions for maximizing the absorption achieved for each coefficient.

In the case of a small plasmonic sphere ($\varepsilon'<0,~\varepsilon''>0$), the electric dipole resonance is dominating, leading to the following expressions for the partial extinction, scattering, and absorption efficiency

\begin{equation}
 \sigma_{ext}=\frac{6}{x^2}\frac{R_R(R_R+R_L)+X_R(X_R+X_L)}{(R_R+R_L)^2+(X_R+X_L)^2}
\end{equation}

\begin{equation}
 \sigma_{sca}=\frac{6}{x^2}\frac{R_R^2+X_R^2}{(R_R+R_L)^2+(X_R+X_L)^2}
\end{equation}

\begin{equation}
 \sigma_{ext}-\sigma_{sca}=\sigma_{abs}=\frac{6}{x^2}\frac{R_RR_L+X_RX_L}{(R_R+R_L)^2+(X_R+X_L)^2}
\end{equation}

Similar to circuit theory, the absorption efficiency is maximized when the partial derivatives with respect to $R_R$ and $X_R$ are simultaneously zero. Alternatively, 
\begin{equation}
 \frac{\partial \sigma_{abs}(R_L,X_L)}{\partial R_L}=0,~\text{and}~\frac{\partial \sigma_{abs}(R_L,X_L)}{\partial X_L}=0
\end{equation}
After some algebra, the above expressions are mutually satisfied when 
\begin{equation}\label{RL_XL_condition}
 R_R=R_L,~\text{and}~X_R=X_L
\end{equation}
Since $Z_R$ is primarily attributed to radiative damping mechanisms ($x^3$) and $Z_L$ incorporate constant (static) and dynamic ($x^2$) terms it is clear that the absorption maximum takes place when the two internal mechanisms are \emph{matched}~\cite{Tretyakov2014}. For $Z_R=Z_L$ both impedances are equivalent and Eq.~(\ref{eq:circuit}) yields to the real value of $a^P_1=\frac{1}{2}$, hence this is the amplitude for which the absorption efficiency is maximized for every coefficient. This value stresses the fact that the absorbed power can be $\propto\left|a^P_1\right|^2=\frac{1}{4}$ on its maximum value~\cite{Liberal2013}. 

By solving the above matching condition we obtain
\begin{equation}\label{Qabs_max_e}
 \varepsilon'=-2-\frac{12}{5}x^2+...
\end{equation}

\begin{equation}\label{Qabs_max_ee}
 \varepsilon''=2x^3+...
\end{equation}
Surprisingly, condition~(\ref{Qabs_max_e}) fits exactly the pole condition (real part of Eq.~(\ref{a1_pole_2})) for the first electric Mie coefficient, while condition~(\ref{Qabs_max_ee}) coincides with the radiative damping term (imaginary part of Eq.~(\ref{a1_pole_2})).

The same analysis can be followed for any multipole term e.g., the absorption maximum occurring for the first magnetic coefficient is approximately at $2x$; the radiative damping mechanism affects not only the width of the resonances but also the amount of losses required for maximum resonant absorption. This fact can be exploited for inverse-scattering purposes, where the determination of the level of losses can be extracted by the level of the absorption efficiency, since the maximum of the latter occurs for losses matching the radiative damping.


\section{Summary}

This article presented the use of Pad\'e approximants as an alternative system ansatz for the Mie coefficients. This perspective provides insights into the scattering mechanisms, i.e., static, dynamic, and radiative damping processes. Additionally, the higher order Pad\'e approximants are used for the extraction of accurate pole condition formulas for every resonance. The accuracy of the proposed formulas is relatively high for size parameters up to $0.5$ for all cases. 

It has been proved that the overall scattering characteristics, such as the resonant linewidth and the absorption maximum, are affected by the radiative damping and dissipative mechanisms. The intrinsic resonant absorption limitations were obtained utilizing a simple, circuit-inspired analogy; the position and strength of the optimum absorption is a manifestation of the matching between the radiative and absorptive processes.

We foresee that a similar, Pad\'e-based analysis can be implemented for various canonical problems, such as cylinders and core-shell spheres, where simple resonant scattering conditions may facilitate a deeper and more physically intuitive interpretation of their response. Similarly, the existence of analytical resonant conditions may also support the reverse material engineering process, where natural or artificial materials can be optimally used for the design of simple or complex light controlling concepts and applications.

\appendices
\section{Simple Pad\'e expansions of spherical Bessel, Hankel, and Riccati--Bessel/Hankel functions}\label{Ap_a}

In this section we give the simple Pad\'e expansions of some necessary functions, such as the spherical Bessel functions and their derivatives. These terms can be readily used for a simple estimation of the $a_1$ and $b_1$ coefficients. Notice that the Maclaurin expansion of each numerator and denominator term of a rational expression is not always equivalent with the Pad\'e expansion of the same function.

\begin{table}[h]
\begin{center}
\caption{Pad\'e approximants of $j_n$ or $h^{(1)}$ and their derivatives}

\begin{tabular}{l}
\hline
\\
$\begin{aligned}[c] 
 &j_1(x)\approx\frac{x}{3\left(1+\frac{1}{10}x^2\right)}\\
 &h_1^{(1)}(x)\approx-i\frac{1-i\frac{2}{3}x+\frac{1}{2}x^2}{x^2+i\frac{2}{3}x^3}\\
\end{aligned}$
$\begin{aligned}
 &j_2(x)\approx\frac{x^2}{15\left(1+\frac{1}{14}x^2\right)}\\
 &h_2^{(1)}(x)\approx\frac{-i-\frac{2x}{3}-i\frac{x^2}{2}}{x^2+i\frac{2}{3}x^3}\\
\end{aligned}$ 
\\
\\
\hline
\\
$\begin{aligned}[c] 
&\left[xj_1(x)\right]'\approx\frac{2x}{3\left(1+\frac{1}{5}x^2\right)}\\
&\left[xh_1^{(1)}(x)\right]'\approx i \frac{1-\frac{4}{3}x-\frac{1}{2}x^2}{x^2-i\frac{4}{3}x^3}\\
\end{aligned}$
$\begin{aligned}
&\left[xj_2(x)\right]'\approx\frac{x^2}{5\left(1+\frac{5}{42}x^2\right)}\\
&\left[xh_2^{(1)}(x)\right]'\approx i\frac{6}{x^3}
\end{aligned}$\\
\\
\hline
\end{tabular}
\end{center}
\end{table}

%
%
%
%

Alternative series expansions for the same functions can be found in classical textbooks (\cite[Ch.7.4]{stratton2007electromagnetic} and~\cite[Ch.9]{jackson1999classical}). In these expansion at least one pole and one zero is visible. Also similar expressions were given by Wiscombe~\cite{Wiscombe1980} as a computationally efficient way of evaluating the Mie coefficients for small size parameters. Note that the spherical Bessel and Riccati--Bessel functions are purely real for the approximated terms, while Hankel and Riccati--Hankel are complex functions. A physical interpretation can be given to the above expressions, since spherical Bessel functions are real and resemble a standing wave behavior, while Hankel functions are generally complex, representing traveling waves~\cite{balanis2012advanced}.

\section{Accurate poles for the $b_n$ and $a_n$ coefficients}\label{Ap_b}

In this section the numerical values obtained by the heuristic method described above will be presented with accuracy up to the $4^{\text{th}}$ decimal digit. The subscript denotes the Mie coefficient for which the pole is extracted, e.g., $b_1$, $b_2$, etc. The superscript denotes the order of the Pad\'e approximant used. Tables~\ref{table_bn} and~\ref{table_an} present the pole conditions of the first five $b_n$, and the first four $a_n$ coefficients, respectively.

\begin{table}[h]
\caption{Pole conditions of the first five $b_n$ coefficients}
\label{table_bn}
\begin{tabular}{l}
\hline
\\
$\begin{aligned}
\varepsilon_{b_1}^{[27/2]}=-2+\left(\frac{\pi}{x}\right)^2&+1.6960x^2-1.1232x^4+...\\
&-i2x(1-0.6960x^2+0.4447x^4)+...
\end{aligned}$
\\
\\
$\begin{aligned}
\varepsilon_{b_2}^{[29/2]}=-\frac{2}{3}+\left(\frac{4.4934}{x}\right)^2&+0.2497x^2+0.1289x^4+...\\
&-i\frac{2}{9}x^3(1-0.2507x^2)+...
\end{aligned}$
\\
\\
$\begin{aligned}
\varepsilon_{b_3}^{[31/2]}=-\frac{2}{5}+\left(\frac{5.7634}{x}\right)^2&-0.0351x^2-0.0119x^4+...\\
&-i\frac{2}{225}x^5(1-0.1560x^2)+...
\end{aligned}$
\\
\\
$\begin{aligned}
\varepsilon_{b_4}^{[33/2]}=-\frac{2}{7}+\left(\frac{6.9879}{x}\right)^2&-0.0119x^2-0.0010x^46+...\\
&-i\frac{2}{11025}x^7(1-0.1166x^2)+...
\end{aligned}$
\\
\\
$\begin{aligned}
\varepsilon_{b_5}^{[35/2]}=-\frac{2}{9}+\left(\frac{8.1827^2}{x^2}\right)^2&-0.0055x^2-0.0002x^4+...\\
&-i\frac{2}{893025}x^9+...
\end{aligned}$
\end{tabular}
\end{table}

\begin{table}[h]
\caption{Pole conditions of the first four $a_n$ coefficients}
\label{table_an}
\begin{tabular}{l}
\hline
\\
$\begin{aligned}
\varepsilon_{a_1}^{[21/2]}=-2+\left(\frac{4.4934}{x}\right)^2&-2.05137x^2+...\\
&-i2x^3(1-\frac{0.0040}{x^2})+...
\end{aligned}$
\\
\\
$\begin{aligned}
\varepsilon_{a_2}^{[23/2]}=-1+\left(\frac{5.7634}{x}\right)^2&-0.1715x^2-0.0858x^4+...\\
&-i\frac{2}{36}x^5(1-\frac{0.0954}{x^2})+...
\end{aligned}$
\\
\\
$\begin{aligned}
\varepsilon_{a_3}^{[27/2]}=-\frac{2}{3}+\left(\frac{6.9879}{x}\right)^2&-0.0463x^2-0.0056x^4+...\\
&-i\frac{2}{2025}x^7(1-\frac{0.3461}{x^2})+...
\end{aligned}$
\\
\\
$\begin{aligned}
\varepsilon_{a_4}^{[29/2]}=-\frac{2}{4}+\left(\frac{8.1825}{x}\right)&-0.0019x^2-0.0011x^4+...\\
&-i\frac{2}{176400}x^9(1-\frac{1.3923}{x^2})+...
\end{aligned}$

\end{tabular}

\end{table}

\section*{Acknowledgment}

The authors would like to thank Dr. Anu Lehtovuori and Prof. Sergei Tretyakov for the stimulating discussions about the circuit analogy of the scattered problem.


\balance
\ifCLASSOPTIONcaptionsoff
  \newpage
\fi



\bibliographystyle{IEEEtran}
\bibliography{IEEEabrv,pade_I}

\begin{thebibliography}{10}
\providecommand{\url}[1]{#1}
\csname url@samestyle\endcsname
\providecommand{\newblock}{\relax}
\providecommand{\bibinfo}[2]{#2}
\providecommand{\BIBentrySTDinterwordspacing}{\spaceskip=0pt\relax}
\providecommand{\BIBentryALTinterwordstretchfactor}{4}
\providecommand{\BIBentryALTinterwordspacing}{\spaceskip=\fontdimen2\font plus
\BIBentryALTinterwordstretchfactor\fontdimen3\font minus
  \fontdimen4\font\relax}
\providecommand{\BIBforeignlanguage}[2]{{%
\expandafter\ifx\csname l@#1\endcsname\relax
\typeout{** WARNING: IEEEtran.bst: No hyphenation pattern has been}%
\typeout{** loaded for the language `#1'. Using the pattern for}%
\typeout{** the default language instead.}%
\else
\language=\csname l@#1\endcsname
\fi
#2}}
\providecommand{\BIBdecl}{\relax}
\BIBdecl

\bibitem{Kuznetsov2016}
A.~I. Kuznetsov, A.~E. Miroshnichenko, M.~L. Brongersma, Y.~S. Kivshar, and
  B.~Lukyanchuk, ``{Optically resonant dielectric nanostructures},''
  \emph{Science}, vol. 354, no. 6314, pp. aag2472--aag2472, 2016.

\bibitem{Shore2015}
R.~A. Shore, ``{Scattering of an electromagnetic linearly polarized plane wave
  by a multilayered sphere: obtaining a computational form of Mie coefficients
  for the scattered field.}'' \emph{IEEE Antennas Propag. Mag.}, vol.~57,
  no.~6, pp. 69--116, Dec 2015.

\bibitem{Fan2014}
X.~Fan, W.~Zheng, and D.~J. Singh, ``{Light scattering and surface plasmons on
  small spherical particles},'' \emph{Light Sci. Appl.}, vol.~3, no.~6, p. 179,
  Jun 2014.

\bibitem{Jahani2016}
S.~Jahani and Z.~Jacob, ``\BIBforeignlanguage{en}{{All-dielectric
  metamaterials}},'' \emph{\BIBforeignlanguage{en}{Nature Nanotechnology}},
  vol.~11, no.~1, pp. 23--36, Jan 2016.

\bibitem{Tribelsky2015}
M.~I. Tribelsky, J.-M. Geffrin, A.~Litman, C.~Eyraud, and F.~Moreno,
  ``\BIBforeignlanguage{en}{{Small dielectric spheres with high refractive
  index as new multifunctional elements for optical devices.}}''
  \emph{\BIBforeignlanguage{en}{Sci. Rep.}}, vol.~5, p. 12288, Jan 2015.

\bibitem{Lannebere2015}
S.~Lanneb{\`{e}}re and M.~G. Silveirinha, ``{Optical meta-atom for localization
  of light with quantized energy},'' \emph{Nat. Commun.}, vol.~6, p. 8766, Oct
  2015.

\bibitem{Lukyanchuk2010}
B.~Luk'yanchuk, N.~I. Zheludev, S.~A. Maier, N.~J. Halas, P.~Nordlander,
  H.~Giessen, and C.~T. Chong, ``{The Fano resonance in plasmonic
  nanostructures and metamaterials.}'' \emph{Nat. Mater.}, vol.~9, no.~9, pp.
  707--715, 2010.

\bibitem{Tribelsky2006}
M.~I. Tribelsky and B.~S. Luk'yanchuk, ``{Anomalous light scattering by small
  particles},'' \emph{Phys. Rev. Lett.}, vol.~97, no.~26, pp. 1--4, 2006.

\bibitem{baker1996pade}
G.~A. Baker and P.~R. Graves-Morris, \emph{Pad{\'e} approximants}.\hskip 1em
  plus 0.5em minus 0.4em\relax Cambridge University Press, 1996, vol.~59.

\bibitem{Tzarouchis2016c}
D.~C. Tzarouchis, P.~Yl{\"{a}}-Oijala, and A.~Sihvola, ``{Unveiling the
  scattering behavior of small spheres},'' \emph{Phys. Rev. B}, vol.~94,
  no.~14, p. 140301(R), Oct 2016.

\bibitem{bohren2008absorption}
C.~F. Bohren and D.~R. Huffman, \emph{{Absorption and scattering of light by
  small particles}}.\hskip 1em plus 0.5em minus 0.4em\relax John Wiley {\&}
  Sons, 2008.

\bibitem{stratton2007electromagnetic}
J.~A. Stratton, \emph{Electromagnetic theory}.\hskip 1em plus 0.5em minus
  0.4em\relax John Wiley \& Sons, 2007.

\bibitem{sihvola2006character}
A.~H. Sihvola, ``{Character of surface plasmons in layered spherical
  structures},'' \emph{Prog. In Electromagn. Res.}, vol.~62, pp. 317--331,
  2006.

\bibitem{Ruppin1975}
R.~Ruppin, ``{Optical properties of small metal spheres},'' \emph{Phys. Rev.
  B}, vol.~11, no.~8, pp. 2871--2876, 1975.

\bibitem{Osipov2015}
A.~V. Osipov and S.~A. Tretyakov, ``{Analytical approximations of low-frequency
  scattering from homogeneous spheres},'' in \emph{2015 1st URSI Atl. Radio
  Sci. Conf. (URSI AT-RASC)}.\hskip 1em plus 0.5em minus 0.4em\relax IEEE, May
  2015.

\bibitem{Wiscombe1980}
W.~J. Wiscombe, ``\BIBforeignlanguage{EN}{{Improved Mie scattering
  algorithms.}}'' \emph{\BIBforeignlanguage{EN}{Appl. Optic.}}, vol.~19, no.~9,
  pp. 1505--9, May 1980.

\bibitem{Lam1992}
C.~C. Lam, P.~T. Leung, and K.~Young, ``{Explicit asymptotic formulas for the
  positions, widths, and strengths of resonances in Mie scattering},'' \emph{J.
  Opt. Soc. Am. B}, vol.~9, no.~9, p. 1585, Sep 1992.

\bibitem{Conwell1984}
P.~R. Conwell, P.~W. Barber, and C.~K. Rushforth, ``{Resonant spectra of
  dielectric spheres},'' \emph{J. Opt. Soc. Am. A}, vol.~1, no.~1, p.~62, Jan
  1984.

\bibitem{Alam2006}
M.~Alam and Y.~Massoud, ``{RLC ladder model for scattering in single metallic
  nanoparticles},'' \emph{IEEE Trans. Nanotechnol.}, vol.~5, no.~5, pp.
  491--498, Sep 2006.

\bibitem{Ambrosio2012}
L.~A. Ambrosio and H.~E. Hernández-Figueroa, ``{RLC circuit model for the
  scattering of light by small negative refractive index spheres},'' \emph{IEEE
  Trans. Nanotechnol.}, vol.~11, no.~6, pp. 1217--1222, Nov 2012.

\bibitem{Mei2013}
Z.~Mei, T.~K. Sarkar, and M.~Salazar-Palma, ``{A study of negative permittivity
  and permeability for small sphere},'' \emph{IEEE Antennas Wirel. Propag.
  Lett.}, vol.~12, pp. 1228--1231, 2013.

\bibitem{Meier1983}
M.~Meier and A.~Wokaun, ``\BIBforeignlanguage{EN}{{Enhanced fields on large
  metal particles: dynamic depolarization}},''
  \emph{\BIBforeignlanguage{EN}{Optic. Lett.}}, vol.~8, no.~11, p. 581, nov
  1983.

\bibitem{DeVries1998}
P.~de~Vries, D.~V. van Coevorden, and A.~Lagendijk, ``{Point scatterers for
  classical waves},'' \emph{Rev. Mod. Phys.}, vol.~70, no.~2, pp. 447--466, Apr
  1998.

\bibitem{Masjuan2014}
P.~Masjuan, J.~{Ruiz de Elvira}, and J.~J. Sanz-Cillero, ``{Precise
  determination of resonance pole parameters through Pad{\'{e}}
  approximants},'' \emph{Phys. Rev. D}, vol.~90, no.~9, p. 097901, Nov 2014.

\bibitem{Liberal2016}
I.~Liberal, A.~M. Mahmoud, and N.~Engheta, ``{Geometry-invariant resonant
  cavities},'' \emph{Nat. Commun.}, vol.~7, p. 10989, Mar 2016.

\bibitem{Tribelsky2011}
M.~I. Tribelsky, ``\BIBforeignlanguage{en}{{Anomalous light absorption by small
  particles}},'' \emph{\BIBforeignlanguage{en}{Europhys. Lett.}}, vol.~94,
  no.~1, p. 14004, Apr 2011.

\bibitem{Johnson1993}
B.~R. Johnson, ``{Theory of morphology-dependent resonances: shape resonances
  and width formulas},'' \emph{J. Opt. Soc. Am. A}, vol.~10, no.~2, p. 343, Feb
  1993.

\bibitem{Averitt1999}
R.~D. Averitt, S.~L. Westcott, and N.~J. Halas, ``{Linear optical properties of
  gold nanoshells},'' \emph{J. Opt. Soc. Am. B}, vol.~16, no.~10, p. 1824, Oct
  1999.

\bibitem{Carminati2006}
R.~Carminati, J.-J. Greffet, C.~Henkel, and J.~Vigoureux, ``{Radiative and
  non-radiative decay of a single molecule close to a metallic nanoparticle},''
  \emph{Opt. Commun.}, vol. 261, no.~2, pp. 368--375, 2006.

\bibitem{Collin1964}
R.~E. Collin and S.~Rothschild, ``{Evaluation of antenna Q},'' \emph{IEEE
  Trans. Antennas Propag.}, vol.~12, no.~1, pp. 23--27, Jan 1964.

\bibitem{jackson1999classical}
J.~D. Jackson, \emph{Classical electrodynamics}.\hskip 1em plus 0.5em minus
  0.4em\relax Wiley, 1999.

\bibitem{Gustafsson2012}
M.~Gustafsson and B.~L.~G. Jonsson, ``{Stored electromagnetic energy and
  antenna Q},'' \emph{Prog. Electromagn. Res.}, vol. 150, no. November 2014,
  pp. 1--34, 2012.

\bibitem{Hansen2012}
T.~V. Hansen, O.~S. Kim, and O.~Breinbjerg, ``{Stored energy and quality factor
  of spherical wave functions-in relation to spherical antennas with material
  cores},'' \emph{IEEE Trans. Antennas Propag.}, vol.~60, no.~3, pp.
  1281--1290, Mar 2012.

\bibitem{Alu2011}
A.~Al{\`{u}}, A.~D. Yaghjian, R.~A. Shore, and M.~G. Silveirinha, ``{Causality
  relations in the homogenization of metamaterials},'' \emph{Phys. Rev. B},
  vol.~84, no.~5, p. 054305, Aug 2011.

\bibitem{Tretyakov2014}
S.~Tretyakov, ``{Maximizing absorption and scattering by dipole particles},''
  \emph{Plasmonics}, pp. 935--944, 2014.

\bibitem{Liberal2013}
I.~Liberal and R.~W. Ziolkowski, ``{Analytical and equivalent circuit models to
  elucidate power balance in scattering problems},'' \emph{IEEE Trans. Antennas
  Propag.}, vol.~61, no.~5, pp. 2714--2726, May 2013.

\bibitem{balanis2012advanced}
C.~A. Balanis, \emph{Advanced Engineering Electromagnetics}, 2012.

\end{thebibliography}

\end{document}